\documentclass[aps,prl,twocolumn,groupedaddress,showpacs,showkeys]{revtex4} 
\usepackage{latexsym,epsfig,amsmath,amssymb}
\usepackage{amssymb}
\usepackage{amsmath}
\usepackage{amsfonts}
\usepackage{epsfig} 
\usepackage{verbatim} 

\newcommand{\seq}{\begin{subequations}}
\newcommand{\sen}{\end{subequations}}
\newcommand{\eq}{\begin{eqnarray}}
\newcommand{\en}{\end{eqnarray}}
\newcommand{\ra}{\rangle}

\def\jpsi{J_\psi}

\begin{document}

\title{Hadronic molecule structure of the $Y(3940)$ and $Y(4140)$} 

\noindent
\author{Tanja Branz, 
        Thomas Gutsche, 
        Valery E. Lyubovitskij\footnote{On leave of absence
        from Department of Physics, Tomsk State University,
        634050 Tomsk, Russia} 
\vspace*{1.2\baselineskip}}

\affiliation{Institut f\"ur Theoretische Physik,  
Universit\"at T\"ubingen,\\
Kepler Center for Astro and Particle Physics, \\ 
Auf der Morgenstelle 14, D--72076 T\"ubingen, Germany
}
    
\date{\today}

\begin{abstract} 

We report on further evidence that the $Y(3940)$ and the recently
observed $Y(4140)$ are heavy hadron molecule states 
with quantum numbers $J^{\rm PC} = 0^{++}$. The $Y(3940)$ state is
considered to be a superposition of 
$D^{\ast +}D^{\ast -}$ and $D^{\ast 0} \overline{D^{\ast 0}}$, 
while  the $Y(4140)$ is a bound state of $D_s^{\ast +}$ and 
$D_s^{\ast -}$ mesons. For the first time we give predictions for the strong
$Y(3940) \to J/\psi \omega$, $Y(4140) \to J/\psi \phi$
and radiative $Y(3940)/Y(4140) \to \gamma\gamma$ decay widths in
a phenomenological Lagrangian approach.
Results for the strong $J/\psi V$ ($V=\omega , \phi$) decays
clearly support the molecular interpretation of the $Y(3940)$
and $Y(4140)$. The alternative assignment of
$J^{\rm PC} = 2^{++}$ is also tested, giving similar results for
the strong decay widths.

\end{abstract}



\maketitle

\newpage

The recent announcement~\cite{Aaltonen:2009tz} of the narrow
state $Y(4140)$ by the CDF Collaboration at Fermilab raises the
prospects of possibly uniquely identifying the structure of a meson resonance 
which does not fit in the conventional quark-antiquark picture.
This latest report of a narrow structure with charmonium-like decay modes
is a continuation of previous discoveries of such states~\cite{Amsler:2008zz} 
which are not easily explained as quark-antiquark configurations.
Possible alternative interpretations involve structures such as
hadronic molecules, tetraquark states or even hybrid configurations
(for recent reviews see e.g. Refs.~\cite{Swanson:2006st,Eichten:2007qx}). 
Now the CDF Collaboration has evidence 
of a narrow near-threshold structure, termed the $Y(4140)$ meson, 
in the $J/\psi \phi$ mass spectrum in exclusive $B^+ \to J/\psi \phi K^+$
decays with the mass $m_{Y(4140)} = 4143.0 \pm 2.9 ({\rm stat}) \pm 
1.2 ({\rm syst})$ MeV 
and natural width $\Gamma_{Y(4140)} = 11.7^{+8.3}_{-5.0} ({\rm stat}) 
\pm 3.7 ({\rm syst})$ MeV~\cite{Aaltonen:2009tz}.
As already stressed in \cite{Aaltonen:2009tz}, the new structure $Y(4140)$,
which decays to $J/\psi \phi$ just above the $J/\psi \phi$ threshold, 
is similar to the previously discovered 
$Y(3940)$~\cite{Abe:2004zs,Aubert:2007vj}, 
which decays to $J/\psi\omega$ near this respective threshold.
The mass and width of the $Y(3940)$ resonance are:
$m_{Y(3940)} = 3943 \pm 11 ({\rm stat}) \pm 13 ({\rm syst})$ MeV,
$\Gamma_{Y(3940)} = 87 \pm 22 ({\rm stat}) \pm 26 ({\rm syst})$ MeV
(Belle Collaboration~\cite{Abe:2004zs}) and
$m_{Y(3940)} = 3914.6^{+3.8}_{-3.4} ({\rm stat}) \pm 2.0 ({\rm syst})$ MeV,
$\Gamma_{Y(3940)} = 34^{+12}_{-8} ({\rm stat}) \pm 5 ({\rm syst})$
MeV ({\it BABAR}~\cite{Aubert:2007vj}).
Both observed states, $Y(4140)$ and $Y(3940)$, are well
above the threshold for open charm decays. A conventional
$c\bar c$ charmonium interpretation is disfavored, since open charm decay
modes would dominate, while the $J/\psi \phi$ or $J/\psi \omega$ decay rates
are essentially negligible~\cite{Aaltonen:2009tz,Eichten:2007qx}. 
Note, current data imply a lower bound of 
$\Gamma(Y(3940) \to J/\psi \omega) > 1$~MeV~\cite{Eichten:2007qx}, which 
is an order of magnitude higher than typical rates between known charmonium 
states. This could be a signal for nonconventional structure of the $Y(3940)$.  
As a first follow-up to the CDF result, it is suggested in~\cite{Liu:2009ei} 
that both the $Y(3940)$ and $Y(4140)$ are hadronic molecules 
(i.e. bound states of mesons induced by the strong interaction). 
These hadron bound states can have quantum numbers $J^{\rm PC} = 0^{++}$ or 
$2^{++}$ whose constituents are the vector charm $D^\ast (D^\ast_s)$ mesons:
\eq\label{M_str} 
|Y(3940)\ra &=& \frac{1}{\sqrt{2}} \big(| D^{\ast +} D^{\ast -} \ra+ 
|D^{\ast 0} \overline{D^{\ast 0}} \ra \big)\,, \nonumber\\ 
|Y(4140)\ra &=& | D^{\ast +}_s D^{\ast -}_s \ra \,. 
\en
The authors of Ref.~\cite{Liu:2009ei} show that binding of the above-mentioned 
meson configurations can be achieved in the context of meson-exchange 
potentials generated by the Lagrangian of heavy hadron chiral perturbation 
theory (HHChPT)~\cite{Wise:1992hn}-\cite{Colangelo:2003sa}. 
Earlier results based on the pion-exchange mechanism already indicated 
that the $D^\ast \overline{D^\ast}$ system can form a bound 
state~\cite{Tornqvist:1993ng}. Binding in the $D^\ast_s \overline{D^\ast_s}$ 
channel can be induced by $\eta$ and $\phi$ meson exchange~\cite{Liu:2009ei}.
A first QCD sum rule study cannot support the claim that the 
$D^\ast_s \overline{D^\ast_s}$ system binds~\cite{Wang:2009ue}.
This issue remains to be studied.

In this Letter we report on a first quantitative prediction for
the decay rates of the observed modes $Y(3940) \to J/\psi \omega$ and
$Y(4140) \to  J/\psi \phi $ assuming the hadronic molecule structures
of Eq.~(\ref{M_str}) with quantum numbers $J^{\rm PC} = 0^{++}$. 
Results will be shown to be fully consistent with present experimental 
observations, strengthening the unusual hadronic molecule interpretation. 
Further predictions are given for the radiative two-photon
decays of these states. Finally, we also consider the alternative
$J^{\rm PC} = 2^{++}$ assignment for the $Y$ states.

The  method of determining the decay rates is based on an effective Lagrangian
which includes both the coupling of the molecular-bound state to their
hadronic constituents and the coupling of the constituents to other hadrons 
and photons. In Refs.~\cite{Faessler:2007gv} we developed the formalism for 
the structural study of other recently observed meson states
(like $D_{s0}^\ast(2317)$, $D_{s1}(2460)$, $X(3872)$, 
$\cdots$) as hadronic molecules. The composite (molecular) structure of 
the $Y(3940)$ and $Y(4140)$ states is defined by the compositeness condition 
$Z=0$~\cite{Weinberg:1962hj}-\cite{Ivanov:1996pz}  
(see also Refs.~\cite{Faessler:2007gv}).  
This condition implies that the renormalization constant 
of the hadron wave function 
is set equal to zero or that the hadron exists as a bound state of its 
constituents. Decay processes are then described by the coupling of the final
state particles via one-loop meson diagrams to the constituents of the 
molecular state (see details in~\cite{Faessler:2007gv}). 

For the observed $Y(3940)$ and $Y(4140)$ states  
we adopt the convention that the spin and parity 
quantum numbers of both states are $J^{\rm PC} = 0^{++}$. 
Presently, except for $C=+$, the $J^{\rm P}$ quantum numbers 
are not unambiguously determined yet in experiment. 
For example, the $Y(3940)$ is also discussed as a $J^{\rm PC} = 1^{++}$ 
charmonium candidate~\cite{Eichten:2007qx}, but $0^{++}$ is not ruled out. 
Their masses are expressed in terms of the binding energy
$\epsilon_Y$ as 
$m_{Y(3940)} = 2 m_{D^\ast} - \epsilon_{Y(3940)}$ and  
$m_{Y(4140)} = 2 m_{D^\ast_s} - \epsilon_{Y(4140)}$, 
where $m_{D^\ast} \equiv m_{D^{\ast \, +}} = 2010.27$ MeV and 
$m_{D^\ast_s} = m_{D^{\ast \, +}_s} = 2112.3$ MeV are the masses 
of the constituent mesons. Since the observed masses are
relatively far from the corresponding 
thresholds we do not include isospin-breaking effects (i.e. 
we suppose that charged and neutral nonstrange $D^\ast$ mesons 
have the same masses). 
Following Ref.~\cite{Liu:2009ei} we consider the $Y(3940)$ 
meson as a superposition of the molecular $D^{\ast +}D^{\ast -}$ and 
$D^{\ast 0}D^{\ast 0}$ states, while the $Y(4140)$ is a bound state of 
$D^{\ast +}_s$ and $D^{\ast -}_s$ mesons (see Eq.~(\ref{M_str})). 
The coupling of the scalar molecular states to their constituents is
expressed by the phenomenological Lagrangian: 
\eq\label{LY1} 
{\cal L}_Y(x) = g_Y Y_{ij}(x) J_{Y_{ij}}(x) 
\en 
where $g_Y$ is the coupling constant; 
$Y_{ij}$ is the $3 \times 3$ matrix containing a nonet of  
possible hidden and open flavor $Y$ states which can be 
composed of vector $D^\ast (D^\ast_s)$ mesons: 
\eq 
Y_{ij} = \left(
\begin{array}{ccc}
\displaystyle{
\frac{Y^0_\rho}{\sqrt{2}} + \frac{Y_\omega}{\sqrt{2}}} 
& Y^+_\rho & Y^+_{K^\ast} \\[2mm] 
Y^-_\rho & - \displaystyle{\frac{Y^0_\rho}{\sqrt{2}} 
+ \frac{Y_\omega}{\sqrt{2}}} & Y^0_{K^\ast} \\[2mm] 
Y^-_{K^\ast} & \overline{Y^0_{K^\ast}} & Y_\phi \\ 
\end{array}
\right) \,. 
\en 
In addition to the detected $Y_\omega = Y(3940)$ 
and $Y_\phi = Y(4140)$ states, one can propose an isotriplet 
of nonstrange states $Y^+_\rho = (D^{\ast +} \overline{D^{\ast 0}})$, 
$Y^-_\rho = (D^{\ast -} D^{\ast 0})$, 
$Y^0_\rho = (D^{\ast 0}\overline{D^{\ast 0}}-D^{\ast +}D^{\ast -})/\sqrt{2}$  
and two isodoublets of strange states  
$Y^+_{K^\ast} = (\overline{D^{\ast 0}} D^{\ast +}_s)$, 
$Y^0_{K^\ast} = (D^{\ast -} D^{\ast +}_s )$ and 
$Y^-_{K^\ast} = (D^{\ast 0} D^{\ast -}_s)$, 
$\overline{Y^0_{K^ \ast}} = (D^{\ast +} D^{\ast -}_s)$. 
We expect that the masses of the $Y^\pm_\rho, Y^0_\rho$ states are 
close to the $Y(3940)$ mass, while the masses of the other four states 
$Y^\pm_{K^\ast}$, $Y^0_{K^\ast}$, $\overline{Y^0_{K^\ast}}$ 
could be approximately 4040 MeV $\simeq$ $m_{D^\ast}$ $+$  
$m_{D^\ast_s}$ $-$ $80$ MeV (a typical value for the binding energy, 
as in the case for the $Y(3940)$ and $Y(4140)$ states). 
In analogy with the $Y(3940)$ and $Y(4140)$ states, we suggest that the new 
hypothetical states $Y^{\pm(0)}_\rho(3940)$ and $Y^{\pm(0)}_{K^\ast}(4040)$ 
can decay into $J/\psi \rho$ and $J/\psi K^\ast$ pairs, respectively. 
Note, we use the notation $Y_V$ for the nonet of $Y$ states since it decays 
into $J/\psi V$ pairs. $J_{Y_{ij}}$ of Eq.~(\ref{LY1}) is the current 
composed of the constituents of the respective hadronic molecule $Y_{ij}$. 
The simplest form of the hadronic currents $J_{Y_{ij}}$ is given by: 
$J_{Y_{ij}}(x) = g_{\mu\nu} \int d^4 y \Phi(y^2) J^{\mu\nu}_{ij}(x,y)$, 
where $J^{\mu\nu}_{ij}(x,y)=D^{\ast\mu}_{i}(x+\frac{y}{2}) 
D^{\ast\nu\dagger}_{j}(x-\frac{y}{2}).$ 
Here $\Phi(y^2)$ is the correlation function describing the 
distribution of the constituents inside the molecular states~$Y$.  
For simplicity we adopt a universal equivalent function for all states.    
A basic requirement for the choice of an explicit form of the correlation 
function $\Phi$ is that its Fourier transform vanishes at a sufficient rate  
in the ultraviolet region of Euclidean space to render the Feynman diagrams 
ultraviolet finite. We use a Gaussian form of $\Phi(y^2)$: 
$\tilde\Phi(p_E^2/\Lambda^2_Y) \doteq \exp( - p_E^2/\Lambda_Y^2)$, 
where $p_{E}$ is the Euclidean Jacobi momentum. Here, $\Lambda_Y$
is a size parameter with a value of about 2 GeV -- a typical 
scale for the masses of the constituents of the $Y$ states.  
The coupling constants $g_Y$ are determined by the compositeness 
condition~\cite{Faessler:2007gv}-\cite{Ivanov:1996pz}: 
$Z_Y = 1 - \Sigma_Y^\prime(m_Y^2) = 0$, where  
$\Sigma^\prime_Y(m_Y^2) = d\Sigma_Y(p^2)/dp^2|_{p^2=m_Y^2}$ 
is the derivative of the mass 
operator $\Sigma_Y$ generated by ${\cal L}_Y(x)$.  

To determine the strong $Y \to J/\psi V$ and two-photon 
$Y \to \gamma\gamma$ decays we have to include the couplings 
of $D^\ast(D^{\ast}_s)$ mesons to vector mesons 
($J/\psi$, $\omega$, $\phi$) and to photons. 
The couplings of $J/\psi$, $\omega$, $\phi$ to vector 
$D^\ast(D^\ast_s)$ mesons are taken from the HHChPT 
Lagrangian~\cite{Wise:1992hn,Colangelo:2003sa}: 
\eq\label{Chiral_Lagrangian}
{\cal L}_{D^\ast D^\ast J_\psi} &=& i g_{_{D^\ast D^\ast J_\psi}}
J_\psi^\mu \Big(
  D^{\ast \dagger}_{\mu i} \,  
\!\stackrel{\leftrightarrow}{\partial}_{\nu}\!D^{\ast \nu}_i
+ D^{\ast \dagger}_{\nu i} \, 
\!\stackrel{\leftrightarrow}{\partial}^{\nu}\!\!D^{\ast}_{\mu i} \nonumber \\
&-& D^{\ast \dagger\nu}_i \,  
\!\stackrel{\leftrightarrow}{\partial}_{\mu}\!D^{\ast}_{\nu i} \Big)
\,, \\
{\cal L}_{D^\ast D^\ast V} &=&  i g_{_{D^\ast D^\ast V}} V^\mu_{ij} 
D^{\ast \dagger}_{\nu i} \, 
\!\stackrel{\leftrightarrow}{\partial}_{\mu}\!D^{\ast \nu}_j \nonumber \\  
&+& 4 i f_{_{D^\ast D^\ast V}} 
(\partial^\mu V^\nu_{ij} - \partial^\nu V^\mu_{ij}) 
D^{\ast}_{\mu i} D^{\ast \dagger \nu}_j \nonumber
\en  
where $A \!\! \stackrel{\leftrightarrow}{\partial} \!\! B
\equiv A \partial B - B \partial A$; $i,j$ are flavor indices; 
$V_{ij} = {\rm diag}\{\omega/\sqrt{2}, \omega/\sqrt{2}, \phi\}$ is 
the diagonal matrix containing $\omega$ and $\phi$ mesons  
(we omit the $\rho$ and $K^\ast$ mesons);  
$D^\ast_i = (D^{\ast 0}, D^{\ast +}, D^{\ast +}_s)$ is the 
triplet of vector $D^\ast$ mesons containing light antiquarks 
$\bar u$, $\bar d$ and $\bar s$, respectively. 
The chiral couplings $g_{_{D^\ast D^\ast J_\psi}}$, 
$g_{_{D^\ast D^\ast V}}$ and $f_{_{D^\ast D^\ast V}}$ 
are fixed as~\cite{Wise:1992hn}-\cite{Colangelo:2003sa}: 
$g_{_{D^\ast D^\ast V}} = \beta g_V/\sqrt{2}$\,,
$f_{_{D^\ast D^\ast V}} = m_{D^\ast}\lambda g_V/\sqrt{2}$\,,
$g_{_{D^\ast D^\ast J_\psi}} = (m_{D^\ast} m_{J_\psi})/(m_D f_{J_\psi})$, 
where $f_{J_\psi} = 416.4$~MeV is the $J/\psi$ leptonic decay constant; 
$g_V \approx  5.8$ and $\beta \approx 0.9$ are fixed using 
vector dominance; the parameter $\lambda = 0.56$~GeV$^{-1}$ is extracted 
by matching HHChPT to lattice QCD and light cone sum rules 
(see details in~\cite{Isola:2003fh}). The leading-order process
relevant for the strong decays $Y(3940) \to J/\psi \omega$ and 
$Y(4140) \to J/\psi \phi$ is the diagram of Fig.1
involving the vector mesons $D^\ast$ or $D^\ast_s$ in the loop. 

The coupling of the charged $D^{\ast \pm}(D^{\ast \pm}_s)$ mesons 
to photons is generated by minimal substitution in the free Lagrangian 
of these mesons. The corresponding electromagnetic Lagrangian reads as: 
\eq 
{\cal L}_{\rm em} &=& e A_\alpha \Big( 
   g^{\alpha\nu} D^{\ast -}_\mu i\partial^\mu D^{\ast +}_\nu 
-  g^{\mu\nu} D^{\ast -}_\mu i\partial^\alpha D^{\ast +}_\nu 
+ {\rm H.c} \Big) \nonumber\\  
&+& e^2 D^{\ast -}_\mu D^{\ast +}_\nu 
\Big( A^\mu A^\nu - g^{\mu\nu} A^\alpha A_\alpha \Big)\,.
\en 
This Lagrangian results in the two relevant diagrams displayed
in Figs.2(a) and 2(b). In order to fulfill electromagnetic gauge 
invariance the strong interaction Lagrangian ${\cal L}_Y$ also 
has to be modified. As outlined in Ref.~\cite{Mandelstam:1962mi} 
and extensively used in Refs.~\cite{Ivanov:1996pz,Faessler:2007gv},  
each charged constituent meson field $H^{\pm}$ in ${\cal L}_Y$ 
is multiplied by the gauge field exponential:
$H^{\pm}(y) \to e^{\mp i e I(y,x,P)} H^{\pm}(y)$, where 
$I(x,y,P) = \int_y^x\! dz_\mu A^\mu(z)$. 
Expanding $e^{\mp i e I(y,x,P)}$ up to second 
order in the electromagnetic field, the two additional 
diagrams of Figs.2(c) and 2(d) are generated, which are necessary 
to guarantee full gauge invariance. 
The contribution of these additional processes is significantly 
suppressed (of the order of a few percent) compared to the leading 
diagram of Fig.2(a).  

The invariant matrix elements of the strong and two-photon transitions 
(when all initial and final particles are on their mass shell)  
are given by: 
\eq 
M_{\mu\nu}(Y \to J/\psi V) &=& g_{\mu\nu} \, g_{_{YJ_\psi V}}  
+ v_{2 \mu} v_{1 \nu} \, f_{_{YJ_\psi V}}  \,, \nonumber\\ 
M_{\mu\nu}(Y \to \gamma \gamma) &=& 
( g_{\mu\nu} q_1 q_2 - q_{2 \mu} q_{1 \nu} ) \, g_{_{Y\gamma\gamma}} \,, 
\en 
where $v_1(q_1)$ and $v_2(q_2)$ are the 4-velocities (momenta) 
of $J_\psi$ and $V$. The effective strong couplings $g_{_{YJ_\psi V}}$ 
and $f_{_{YJ_\psi V}}$ have dimension of mass, while the electromagnetic 
coupling $g_{_{Y\gamma\gamma}}$ has dimension of inverse mass. 
The matrix element of the two-photon transition has 
a full gauge-invariant structure. 
The constants $g_{_{YJ_\psi V}}$ and $f_{_{YJ_\psi V}}$ are 
products of the coupling $g_Y$, the chiral couplings in 
Eq.~(\ref{Chiral_Lagrangian}) and the generic $D^\ast$ meson loop 
structure integral (see Fig.1).  
In terms of these effective couplings $g_{_{YJ_\psi V}}$, $f_{_{YJ_\psi V}}$ 
and $g_{_{Y\gamma\gamma}}$ the corresponding decay widths are calculated 
according to the expressions: 
\eq 
\hspace*{-.1cm} 
\Gamma(Y \to J/\psi V) &=& 
\frac{3 P^\ast}{8\pi m_Y^2} g_{_{YJ_\psi V}}^2  
 (1 + \beta + 2 w r \beta + 3 r^2 \beta^2 )\,,   \nonumber\\ 
\hspace*{-.1cm} 
\Gamma(Y \to \gamma\gamma) &=& \frac{\pi}{4} \alpha^2 m_Y^3 
g_{_{Y\gamma\gamma}}^2 \,, 
\en
where 
\eq 
r = \frac{f_{_{YJ_\psi V}}}{g_{_{YJ_\psi V}}}\,, \ \ 
\beta = \frac{1}{3} \, 
\biggl(\frac{P^{\ast}m_Y}{m_{\jpsi} m_V}\biggr)^2\,, \ \ 
w = v_1 v_2 \nonumber 
\en 
and $\alpha$ is the fine structure constant. 
Here $P^\ast$ is the corresponding three-momentum 
of the decay products. 

\begin{figure}
\centering{\
\epsfig{figure=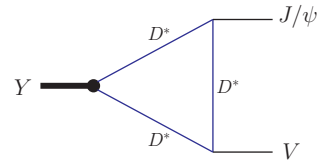,scale=.5}}
\caption{The diagram describes the $Y \to J/\psi V$ decay.}
\label{fig:str}
\end{figure}

\begin{figure} 
\centering{\
\epsfig{figure=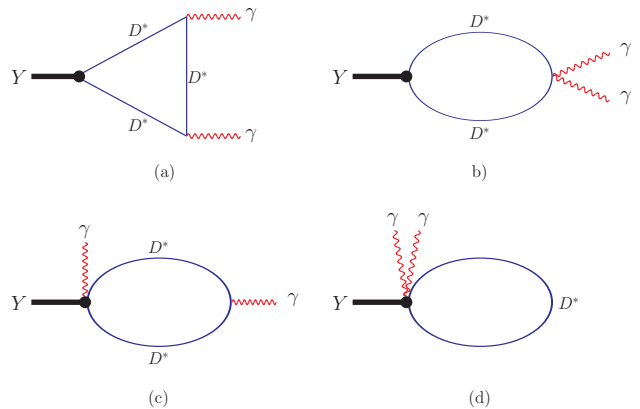,scale=.45}}
\caption{The diagrams shows contributions to the 
$Y \to \gamma\gamma$ decay.}
\label{fig:vertex}
\end{figure}

\begin{table}
\caption{\label{tab:1}Decay properties of $Y(3940)$ and $Y(4140)$ states}  
\begin{ruledtabular}
\begin{tabular}{|l|c|c|}
Quantity                            & $Y(3940)$ & $Y(4140)$ \\ 
\hline 
$g_Y$, GeV                          & 14.08 $\pm$ 0.30 & 13.20 $\pm$ 0.26 \\ 
\hline 
$g_{_{YJ_\psi V}}$, GeV             & 1.72 $\pm$ 0.03 & 1.46 $\pm$ 0.03 \\ 
\hline 
$f_{_{YJ_\psi V}}$, GeV             & 1.64 $\pm$ 0.01 & 1.84 $\pm$ 0.01 \\
\hline 
$\Gamma(Y \to J/\psi V)$, MeV       & 5.47 $\pm$ 0.34 & 3.26 $\pm$ 0.21 \\
\hline
$g_{_{Y\gamma\gamma}} \times 10^2$, GeV$^{-1}$  
& $1.15 \pm 0.01$  
& $1.46 \pm 0.01$ \\
\hline  
$\Gamma(Y \to \gamma\gamma)$, keV   & 0.33 $\pm$ 0.01 & 0.63 $\pm$ 0.01 \\
\hline 
$\displaystyle{R = \frac{\Gamma(Y \to \gamma\gamma)}
{\Gamma(Y \to J/\psi V)}} \times 10^4$   
& $0.61 \pm 0.06$ & $1.93 \pm 0.16$ \\
\end{tabular}
\end{ruledtabular}
\end{table} 

Our numerical results for the quantities characterizing the strong
$J/\psi V$ ($V = \omega, \phi$) and radiative two-gamma decays
of $Y(3940)$ and $Y(4140)$ are contained in Table~\ref{tab:1}. 
For the masses of the $Y$ states we use the values extracted by 
the {\it BABAR}~\cite{Aubert:2007vj} and the CDF~\cite{Aaltonen:2009tz} 
Collaborations. The error bars correspond to the ones of the experimental 
mass values of the $Y$ states. 

The predictions for the couplings $g_Y$ of the $Y$ states to their 
meson constituents are consistent with a trivial estimate using
the Weinberg formula. It was originally derived for the
deuteron as based on the compositeness condition~\cite{Weinberg:1962hj} 
with $g_Y^W = \sqrt{32 \pi} \, m_{D^\ast}^{3/4} \, \epsilon_Y^{1/4}$.
This formula represents the leading term of an expansion in powers of the
binding energy $\epsilon$. Note that this expression can be obtained
in the local limit (i.e. the vertex function approaches the limit
$\Phi(y^2) \to \delta^4(y)$) and when the longitudinal part
$k^\mu k^\nu/m_{D^\ast}^2$ of the constituent vector meson propagator 
is neglected. The numerical results for $g_{Y(3940)}^W = 9.16$ GeV and 
$g_{Y(4140)}^W = 8.91$ GeV are in good agreement with nonlocal results
of $g_{Y(3940)} = 14.08$~GeV and $g_{Y(4140)} = 13.20$~GeV. 

The predictions of $\Gamma( Y(3940) \to J/\psi \omega )=5.47$~MeV
and $\Gamma (Y(4140) \to J/\psi \phi )=3.26$~MeV for the observed
decay modes are sizable and fully consistent with the
upper limits set by present data on the total widths. 
The result for $\Gamma( Y(3940) \to J/\psi \omega )$ is also 
consistent with the lower limit of about 1 MeV~\cite{Eichten:2007qx}.   
Values of a few MeV for these decay widths naturally arise in the
hadronic molecule interpretation of the $Y(3940)$ and $Y(4140)$,
whereas in a conventional charmonium interpretation the $J/\psi V$
decays are strongly suppressed by the Okubo, Zweig and Iizuka 
rule~\cite{Eichten:2007qx}. In addition to the possibility of binding 
the $D^\ast \bar {D^\ast}$ and $D_s^{\ast +} D_s^{\ast -}$
systems~\cite{Liu:2009ei}, present results on the $J/\psi V$ decays give
further strong support to the interpretation of the $Y$ states as 
heavy hadron molecules. 
Further tests of the presented scenario concern the two-photon decay
widths, which we predict to be of the order of 1 keV.

Finally we also test the $J^{\rm PC} = 2^{++}$
assignment not yet ruled out experimentally. The coupling of the molecular
tensor field $Y_{\mu\nu; ij}$ to the two-meson constituent current
$J^{\mu\nu}_{ij}$ is set up as
\eq\label{LYT} 
{\cal L}_{Y_T}(x) = g_{Y_T} Y_{\mu\nu; ij}(x) 
\int d^4 y \, \Phi(y^2) \, J^{\mu\nu}_{ij}(x,y) \,. 
\en 
Proceeding as outlined before we obtain
\eq 
& &\Gamma(Y(3940) \to J/\psi \omega) = 7.48 \pm 0.27 \ {\rm MeV}\,, \nonumber\\
& &\Gamma(Y(4140) \to J/\psi \phi)   = 4.41 \pm 0.16 \ {\rm MeV}\,, \nonumber\\
& &\Gamma(Y(3940) \to \gamma\gamma)  = 0.27 \pm 0.01 \ {\rm keV}\,, \nonumber\\
& &\Gamma(Y(4140) \to \gamma\gamma)  = 0.50 \pm 0.01 \ {\rm keV}\,. 
\en
Since the results for the strong $J/\psi $ decays are quite similar to the
$0^{++}$ case, a $2^{++}$ scenario cannot be ruled out and is also
consistent within a molecular interpretation of the $Y$ states.

A full interpretation of the $Y(3940)$ and $Y(4140)$ states requires: 
i) an experimental determination of the $J^{\rm PC}$ quantum numbers, 
ii) a consistent and hopefully converging study of binding mechanisms in 
the $D^\ast_{(s)} \overline{D^\ast_{(s)}}$ systems and iii) theory and 
experiment to consider the open charm decay
modes, such as $D\bar D$, $D \bar D^\ast$, $D \bar D^\ast \gamma$, etc.,
which are also naturally fed in a charmonium picture.
Ultimately, only a full understanding of the decay patterns of the
$Y(3940)$ and $Y(4140)$ can lead to a unique structure interpretation,
yet present results clearly support the notion of the establishment 
of hadronic molecules in the meson spectrum. 

After submission of this manuscript, calculations both in the potential 
model approach~\cite{Ding:2009vd} and 
in QCD sum rules~\cite{Albuquerque:2009ak} were presented,
which support the original claim that the $D^\ast_s \overline{D^\ast_s}$ 
system binds for $J^{PC}=0^{++}$, hence give further support to the 
interpretation presented here.

\begin{acknowledgments}

This work was supported by the DFG under Contract No. FA67/31-1,
No. FA67/31-2, and No. GRK683. This research is also part of the  
European Community-Research Infrastructure Integrating Activity  
"Study of Strongly Interacting Matter" (HadronPhysics2,  
Grant Agreement No. 227431) and of the President grant 
of Russia "Scientific Schools"  No. 871.2008.2. 

\end{acknowledgments}


\vspace*{-.7cm}

\begin{thebibliography}{99} 

\bibitem{Aaltonen:2009tz}
  T.~Aaltonen {\em et al.}  (The CDF collaboration), 
  Phys.\ Rev.\ Lett.\  {\bf 102}, 242002 (2009). 
\bibitem{Amsler:2008zz}
  C.~Amsler {\em et~al.} (Particle Data Group), 
  Phys. \ Lett. \ B {\bf 667}, 1 (2008).
\bibitem{Swanson:2006st}
  E.~S.~Swanson,
  Phys.\ Rept.\  {\bf 429}, 243 (2006); \\{}
  M.~B.~Voloshin,
  Prog.\ Part.\ Nucl.\ Phys.\  {\bf 61}, 455 (2008).  
\bibitem{Eichten:2007qx}
  E.~Eichten, S.~Godfrey, H.~Mahlke and J.~L.~Rosner,
  Rev.\ Mod.\ Phys.\  {\bf 80}, 1161 (2008); 
  S.~Godfrey and S.~L.~Olsen,
  Ann.\ Rev.\ Nucl.\ Part.\ Sci.\  {\bf 58}, 51 (2008). 
\bibitem{Abe:2004zs}
  S.-K.~Cho {\em et~al.}  (Belle Collaboration),
  Phys.\ Rev.\ Lett.\  {\bf 94}, 182002 (2005). 
\bibitem{Aubert:2007vj}
  B.~Aubert {\em et~al.}  ({\it BABAR} Collaboration),
  Phys.\ Rev.\ Lett.\  {\bf 101}, 082001 (2008). 
\bibitem{Liu:2009ei}
  X.~Liu and S.~L.~Zhu,
  Phys.\ Rev.\  D {\bf 80}, 017502 (2009). 
\bibitem{Wise:1992hn}
  M.~B.~Wise,
  Phys.\ Rev.\  D {\bf 45}, R2188 (1992); 
  T.~M.~Yan, H.~Y.~Cheng, C.~Y.~Cheung, G.~L.~Lin, Y.~C.~Lin and H.~L.~Yu,
  Phys.\ Rev.\  D {\bf 46}, 1148 (1992)
  [Erratum-ibid.\  D {\bf 55}, 5851 (1997)]; 
  R.~Casalbuoni, A.~Deandrea, N.~Di Bartolomeo, 
  R.~Gatto, F.~Feruglio and G.~Nardulli,
  Phys.\ Rept.\  {\bf 281}, 145 (1997).  
\bibitem{Isola:2003fh}
  C.~Isola, M.~Ladisa, G.~Nardulli and P.~Santorelli,
  Phys.\ Rev.\  D {\bf 68}, 114001 (2003)
\bibitem{Colangelo:2003sa}
  P.~Colangelo, F.~De Fazio and T.~N.~Pham,
  Phys.\ Rev.\  D {\bf 69}, 054023 (2004). 
\bibitem{Tornqvist:1993ng}
  N.~A.~Tornqvist,
  Z.\ Phys.\  C {\bf 61}, 525 (1994). 
\bibitem{Wang:2009ue}
  Z.~G.~Wang,
  arXiv:0903.5200 [hep-ph].
\bibitem{Faessler:2007gv}
  A.~Faessler, T.~Gutsche, V.~E.~Lyubovitskij and Y.~L.~Ma,
  Phys.\ Rev.\  D {\bf 76}, 014005 (2007); 
  Phys.\ Rev.\  D {\bf 76}, 114008 (2007); 
  Y.~B.~Dong, A.~Faessler, T.~Gutsche and V.~E.~Lyubovitskij,
  Phys.\ Rev.\  D {\bf 77}, 094013 (2008); 
  T.~Branz, T.~Gutsche and V.~E.~Lyubovitskij,
  Phys.\ Rev.\  D {\bf 79}, 014035 (2009).  
\bibitem{Weinberg:1962hj}
  S.~Weinberg,
  Phys.\ Rev.\  {\bf 130}, 776 (1963). 
\bibitem{Salam:1962ap}
  A.~Salam,
  Nuovo Cim.\  {\bf 25}, 224 (1962). 
\bibitem{Efimov:1993ei}
  G.~V.~Efimov and M.~A.~Ivanov,
  {\it The Quark Confinement Model of Hadrons},
  (IOP Publishing, Bristol $\&$ Philadelphia, 1993). 
\bibitem{Ivanov:1996pz}
  M.~A.~Ivanov, M.~P.~Locher and V.~E.~Lyubovitskij,
  Few Body Syst.\  {\bf 21}, 131 (1996); 
  M.~A.~Ivanov, V.~E.~Lyubovitskij, J.~G.~K\"orner and P.~Kroll,
  Phys.\ Rev.\ D {\bf 56}, 348 (1997); 
  A.~Faessler, T.~Gutsche, B.~R.~Holstein, M.~A.~Ivanov, 
  J.~G.~Korner and V.~E.~Lyubovitskij,
  Phys.\ Rev.\  D {\bf 78}, 094005 (2008). 
\bibitem{Mandelstam:1962mi}
S.~Mandelstam,
Annals Phys.\  {\bf 19}, 1 (1962). 
\bibitem{Ding:2009vd}
  G.~J.~Ding,
  arXiv:0904.1782 [hep-ph].
\bibitem{Albuquerque:2009ak}
  R.~M.~Albuquerque, M.~E.~Bracco and M.~Nielsen,
  Phys.\ Lett.\  B {\bf 678}, 186 (2009); 
  J.~R.~Zhang and M.~Q.~Huang,
  arXiv:0905.4178 [hep-ph].

\end{thebibliography}
\end{document}